\documentclass[  reprint,  amsmath,amssymb,  aps, pra]{revtex4-2}
\usepackage{graphicx}% Include figure files
\usepackage{dcolumn}% Align table columns on decimal point
\usepackage{bm}% bold math
\usepackage[utf8]{inputenc}
\usepackage[T1]{fontenc}
\usepackage{physics}
\usepackage{amsmath}
\usepackage{placeins}
\usepackage{color}
\usepackage[most]{tcolorbox}
\usepackage{tikz}
\usepackage{relsize}
\usepackage[normalem]{ulem}
\usepackage{parskip}
\usepackage{xr}
\usepackage{lineno}
\begin{document}
%\linenumbers
%\preprint{APS/123-QED}

\title{Non-reciprocal Binary-fluid Turbulence}

\author{Biswajit Maji\textsuperscript{1}}
\email{biswajitmaji@iisc.ac.in}

\author{Nadia Bihari Padhan\textsuperscript{2}}
\email{nadia_bihari.padhan@tu-dresden.de}

\author{Axel Voigt\textsuperscript{2,3,4}}
\email{axel.voigt@tu-dresden.de}

\author{Rahul Pandit\textsuperscript{1}}
\email{rahul@iisc.ac.in}

\affiliation{
\textsuperscript{1}Centre for Condensed Matter Theory, Department of Physics, Indian Institute of Science, Bangalore 560012, India \\
\textsuperscript{2}Institute of Scientific Computing, TU Dresden, 01069 Dresden, Germany\\
\textsuperscript{3}Center of Systems Biology Dresden (CSBD), Pfotenhauerstr. 108, 01307 Dresden, Germany \\
\textsuperscript{4} Cluster of Excellence, Physics of Life (PoL), TU Dresden, Arnoldstr. 18, 01307 Dresden, Germany
}

%\thanks{\textsuperscript{$\dagger$}These authors contributed equally.}
\begin{abstract}
    Although effective non-reciprocal interactions have been investigated in a variety of fields, their consequences have not been explored in hydrodynamical turbulence. We initiate such an exploration by introducing non-reciprocal binary-fluid tubulence and uncover its properties by developing a two-dimensional (2D) Non-Reciprocal Cahn-Hilliard-Navier-Stokes (NRCHNS) model. We show that, as we increase the strength of the non-reciprocal terms, this model displays a hitherto unanticipated type of turbulence, with an inverse cascade of energy and an energy spectrum $E(k)\sim k^{-5/3}$, reminiscent of the well-known inverse cascade in forced, 2D fluid turbulence, but distinct from it, in so far as it develops a non-reciprocal flux $\bm J$.  We demonstrate how NRCHNS turbulence  suppresses $J(t) = |\bm J|$, as the Reynolds number increases. We compare and contrast 2D NRCHNS turbulence with its fluid-turbulence counterpart by examining spectra, fluxes, spectral balances, flow topologies, and signatures of multifractality.
\end{abstract}

\date{\today}% 
%\begin{document}

\maketitle

 %%%%%%%%%%%%%%%%%%%%%%%%%%%%%%%%%%%
\begin{figure*}[htp]
  \includegraphics[width=1.0\textwidth]{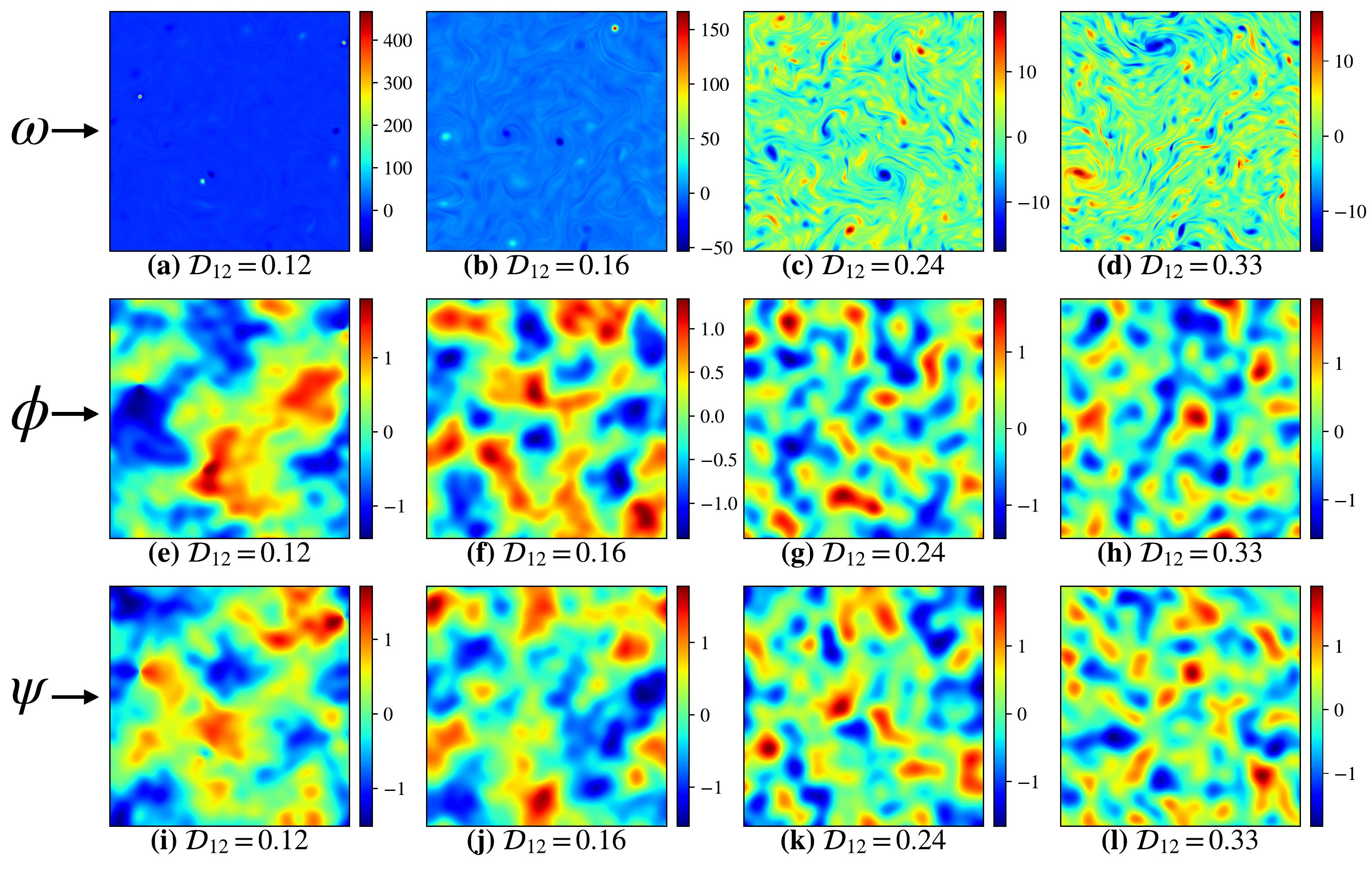}  
\caption{\textcolor{black}{\textbf{Pseudocolor plots of fields in 2D NRCHNS turbulence  at a representative time and with $\nu=0.0001$ and $\sigma_1=\sigma_2=3$:} The non-reciprocity parameter $\mathcal{D}_{12}=0.12$ (column 1), $\mathcal{D}_{12}=0.16$ (column 2), $\mathcal{D}_{12}=0.24$ (column 3), and $\mathcal{D}_{12}=0.33$ (column 4): (a)-(d) $\omega$, with corresponding pseudocolor plots for (e)-(h) $\phi$ and (i)-(l) $\psi$, respectively. For the full spatiotemporal evolution of these fields see the Supplementary Movies V1-V4.}}
\label{fig:pcolor}
\end{figure*}
%%%%%%%%%%%%%%%%%%%%%%%%%%%%%%
%\section{Introduction}
\section{Non-reciprocal Turbulence} 
\label{sec:NRT}
Turbulence continues to be one of the most challenging problems in classical physics, in general, and nonequilibrium statistical mechanics, in particular~\cite{frisch1995turbulence,falkovich2006lessons,sreenivasan2025turbulence}. Classical fluid turbulence is fuelled by external forcing. In statistically steady turbulence, this forcing is balanced by viscous dissipation and the resulting nonequilibrium statistically steady state (NESS) shows remarkable scaling properties for energy spectra $E(k)$, which are spread over several decades of wavenumber $k$, and multifractal velocity and energy-dissipation fluctuations~\cite{frisch1985singularity,meneveau1991multifractal,frisch1995turbulence,boffetta2008twenty}, at large Reynolds numbers $Re$. 
Over the past decade or so, it has become clear that nonequilibrium chaotic flows can also arise when energy is injected \emph{internally}, as in active systems [see, e.g., Ref.~\cite{bowick2022symmetry}], leading to active-fluid turbulence in a variety of systems [see, e.g., Refs.~\cite{alert2022active,pandit2025particles}] that include bacterial suspensions~\cite{sokolov2007concentration,wensink2012meso,dunkel2013minimal,thampi2016active,bratanov2015new}.

In systems that are not in equilibrium, effective non-reciprocity of interactions occurs commonly~\cite{sompolinsky1986temporal,hong2011kuramoto,montbrio2018kuramoto,menzel2013traveling,ivlev2015statistical,kryuchkov2018dissipative,you2020nonreciprocity,gompper20202020,saha2020scalar,hosaka2021nonreciprocal,yasuda2021nonreciprocality,bowick2022symmetry,gupta2022active,poncet2022soft,clerk2022introduction,dinelli2023non,frohoff2023non,frohoff2023nonreciprocal,ryu2023dynamics,tucci2024nonreciprocal,rana2024defect,markovich2024nonreciprocity,mandal2024molecular,reisenbauer2024non,guislain2024collective,al2025non,garces2025phase,avni2025nonreciprocal,pisegna2025can,weis2025generalized,te2025metareview,sahoo2025nonreciprocal,klamser2025directed}. The exploration of such non-reciprocal interactions, in both experimental and model systems, is an exciting and rapidly growing research frontier, which encompasses, \textit{inter alia}, soft and active matter~\cite{gompper20202020,poncet2022soft,gupta2022active,bowick2022symmetry,dinelli2023non,te2023microscopic,markovich2024nonreciprocity,te2025metareview,al2025non,klamser2025directed}, neuronal networks and excitation-inhibition models~\cite{sompolinsky1986temporal,hong2011kuramoto,montbrio2018kuramoto}, phase transitions~\cite{kryuchkov2018dissipative,fruchart2021non}, biological~\cite{halatek2018self} and chemical~\cite{soto2014self,nasouri2020exact,yasuda2021nonreciprocality,mandal2024molecular,liu2024self} systems, and non-Hermitian models~\cite{miri2019exceptional,clerk2022introduction,guislain2024collective}. These studies have yielded a rich variety of results such as pattern formation and traveling waves~\cite{menzel2013traveling,you2020nonreciprocity,saha2020scalar,frohoff2023non,frohoff2023nonreciprocal,brauns2024nonreciprocal} induced by non-reciprocity. Surprisingly, there has been no exploration of non-reciprocal turbulence in any hydrodynamical system. How might we find such turbulence? Our study provides the first answer to this central question by constructing a minimal model in which non-reciprocality is encoded in off-diagonal elements of the diffusion tensor in a binary-fluid mixture. We demonstrate that non-reciprocal turbulence can be induced either by increasing the strength of the non-reciprocal interactions or by reducing the kinematic viscosity $\nu$.
This non-reciprocal turbulence is qualitatively different from the types of turbulence mentioned above: in particular, it does not rely on energy injection via orientational order, force-dipole stresses, or externally imposed body-force injection. Instead, non-reciprocity injects energy through \textit{transport}—localized at composition gradients—so the interface itself becomes the engine that drives turbulence.
\\ \\
To set the stage for our work, we recall that the spatiotemporal evolution of binary mixtures is governed by the Cahn-Hilliard (CH) equations [see, e.g., Ref.~\cite{puri2009kinetics}]. The CH equations employ a scalar order-parameter field, say $\varphi$, to distinguish between two coexisting phases, one rich in component $A$ and another rich in component $B$; this field $\varphi$ assumes positive (negative) values in $A$-rich ($B$-rich) regions. If the two components of the mixture are fluids, then we must couple the CH and Navier-Stokes (NS) equations~\cite{padhan2025cahn}. Furthermore, if there is a source of activity, we must incorporate active-fluid terms, e.g., as in Refs.~\cite{padhan2023activity,padhan2024novel,padhan2025cahn}. The specific type of activity we consider is engendered by non-reciprocity, for which it is natural to use a Non-Reciprocal Cahn-Hilliard-Navier-Stokes (NRCHNS) model [Sec.~\ref{sec:ModMeth}], some versions of which have been used for low-Reynolds-number flows~\cite{menzel2013traveling,you2020nonreciprocity,saha2020scalar,frohoff2023non,frohoff2023nonreciprocal,brauns2024nonreciprocal,blom2026dynamic}. The specific two-dimensional (2D) NRCHNS system we use is:
\begin{eqnarray}
   \pdv{\phi}{t} + \bm{u}\cdot\grad\phi & = & M_1\grad^2[ - \frac{3}{2}\sigma_1\epsilon_1\grad^2\phi \nonumber \\
    &+& \frac{3}{4}\frac{\sigma_1}{\epsilon_1}(-\phi + \phi^3) + D_{12}\psi] \,;\label{eq:phi} \\
\pdv{\psi}{t} + \bm{u}\cdot\grad\psi &=& M_2\grad^2[- \frac{3}{2}\sigma_2\epsilon_2\grad^2\psi \nonumber \\
&+& \frac{3}{4}\frac{\sigma_2}{\epsilon_2}(-\psi + \psi^3) + D_{21}\phi] \,;\label{eq:psi} \\
\pdv{\omega}{t} + \bm{u}\cdot\grad\omega &=& \nu\grad^2\omega  - \frac{3}{2}\sigma_1\epsilon_1\curl{(\grad^2\phi\grad\phi)}\nonumber \\  
&-& \frac{3}{2}\sigma_2\epsilon_2\curl{(\grad^2\psi\grad\psi)}   - 
 \alpha\omega\,;\label{eq:omega}\\
\nabla \cdot \bm u &=& 0\,;\label{eq:incom} 
\end{eqnarray}
the scalar order-parameter fields, $\phi$ and $\psi$ for two Cahn-Hilliard (CH) systems~\cite{padhan2025cahn}, are coupled to an incompressible two-dimensional (2D) Navier-Stokes fluid with kinematic viscosity $\nu$, velocity $\bm u$, and the vorticity $\bm \omega \equiv (\nabla \times \bm u) = \omega \hat{z}$ that points out of the $xy$ plane in the $\hat{z}$ direction; $\alpha$ is the bottom friction; $\sigma_1$ and $\sigma_2$ are surface-tension coefficients; and $\epsilon_1$ and $\epsilon_2$ are the widths for the $\phi$ and $\psi$ interfaces, respectively. For simplicity we set $\alpha = 0$, $\sigma_1=\sigma_2$,  $\epsilon_1=\epsilon_2$,
and the mobilities $M_1=M_2$. We identify the coefficients $-M_1\frac{3}{4}\frac{\sigma_1}{\epsilon_1}$ and $-M_2\frac{3}{4}\frac{\sigma_2}{\epsilon_2}$ as $D_{11}$ and $D_{22}$, the diffusivities for the scalar fields $\phi$ and $\psi$ [see, e.g., Refs.~\cite{saha2020scalar,frohoff2023non,frohoff2023nonreciprocal,brauns2024nonreciprocal}]; their negative signs lead to the well-known CH instability at the linear level, which is then controlled by the nonlinear terms. We take $D_{21}= -D_{12}$ as the off-diagonal elements of the diffusion tensor, which suffices to obtain non-reciprocal binary-fluid turbulence~\footnote{Reference~\cite{brauns2024nonreciprocal} calls this the anti-reciprocal limit.}; and the ratio $\mathcal{D}_{12}\equiv |D_{12}|/|D_{11}|$ provides a natural measure of the non-reciprocity. Note that the terms with the coefficients $D_{12}$ and $D_{21}$, which are responsible for the non-reciprocal nature of this model, do not follow from a functional derivative of a free-energy functional $\mathcal{F}$ [Sec.~\ref{sec:ModMeth}].
%%%%%%%%%%%%%%%%%%%%%%%%%%%%%%
\begin{figure*}[htp]
  \includegraphics[width=1.0\textwidth]{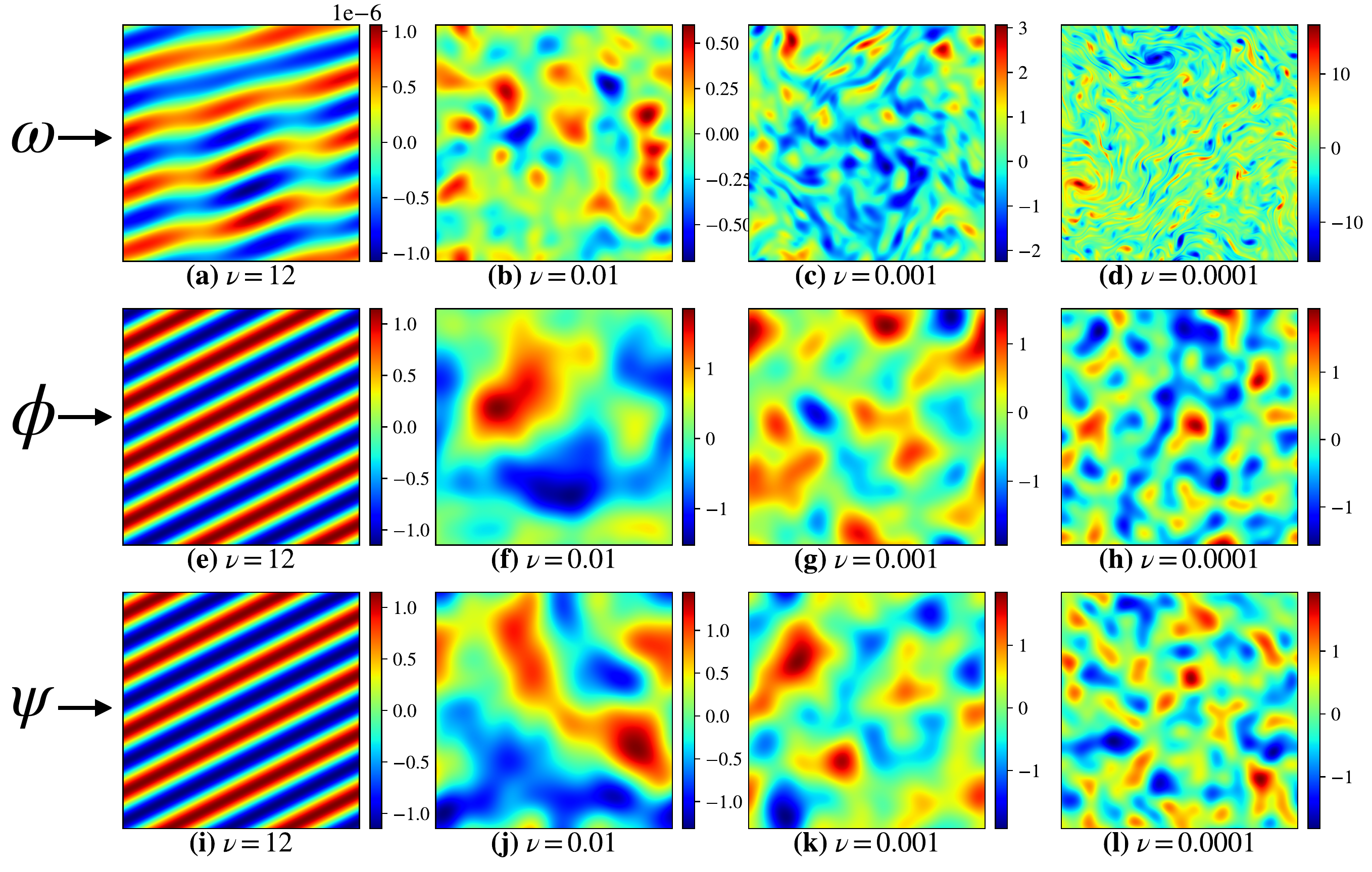}  
\caption{\textcolor{black}{\textbf{Pseudocolor plots for 2D NRCHNS turbulence  at a representative time with the non-reciprocity parameter $\mathcal{D}_{12}=0.33$ and $\sigma_1=\sigma_2=3$:} The kinematic viscosity $\nu=12$ (column 1), $\nu=0.01$ (column 2), $\nu=0.001$ (column 3), and $\nu=0.0001$ (column 4): (a)-(d) $\omega$, with corresponding pseudocolor plots for (e)-(h) $\phi$ and (i)-(l) $\psi$, respectively. For the full spatiotemporal evolution of these fields see the Supplementary Movies V5-V7 and V4.}}
\label{fig:subcolor}
\end{figure*}
%%%%%%%%%%%%%%%%%%%%%%%%%%%%%%%%
 
 Pseudospectral direct numerical simulations (DNSs) [Sec.~\ref{sec:ModMeth}] allow us to study the NRCHNS model in great detail and to obtain its nonequilibrium states. At low $Re$, we find travelling waves that have been explored in low-$Re$ studies~\cite{saha2020scalar,frohoff2023non,frohoff2023nonreciprocal,brauns2024nonreciprocal}. As we increase the strength of the non-reciprocal terms, this model displays a hitherto unanticipated turbulence, with an inverse cascade of energy and an energy spectrum $E(k)\sim k^{-5/3}$, which is reminiscent of the well-known inverse cascade in forced, 2D fluid turbulence~\cite{boffetta2012two,pandit2017overview}. However, there are important differences between 2D fluid and NRCHNS turbulence, which can be characterised by the magnitude $J(t)$ of the non-reciprocal flux~\cite{rana2024defect}
\begin{eqnarray}
\bm{J} \equiv \langle \phi \nabla \psi - \psi \nabla \phi \rangle_s\,,
\label{eq:fluxJ}
\end{eqnarray}
where the subscript $s$ indicates the spatial average. $J(t)$ achieves a steady, finite value at low $Re$; as we lower $\nu$, $Re$ and the intensity of NRCHNS turbulence increase, but $J(t)$ is suppressed. We examine the intricate interplay between non-reciprocity and turbulence and compare and contrast 2D NRCHNS turbulence with its fluid-turbulence counterpart.
%%%%%%%%%%%%%%%%%%%%%%%%%%%%%%%%%
\begin{figure*}[htp]
\centering
\includegraphics[width=1.0\textwidth]{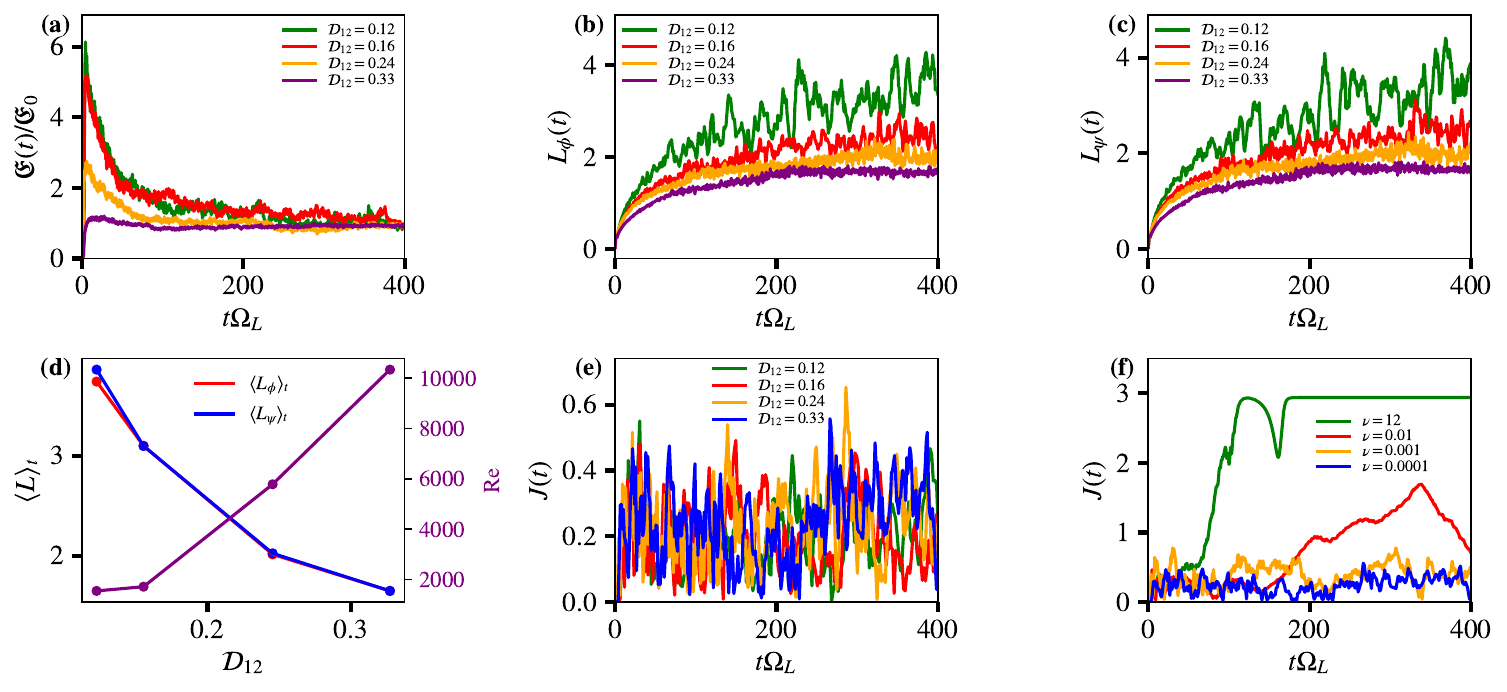}
    \caption{\textbf{Temporal evolution of NRCHNS turbulence:} Plots versus the scaled time $t\Omega_L$ showing
(a) the normalised kinetic energy $\mathfrak{E}(t)/\mathfrak{E}_0$, where $\mathfrak{E}_0 \equiv \mathfrak{E}(t=\Omega_L^{-1})$;
(b) the  coarsening length scale of the composition field, $L_\phi(t)$; and (c) the corresponding length scale of the $\psi$ field, $L_\psi(t)$, for $\mathcal{D}_{12} = 0.12, 0.16, 0.24,$ and $0.33$.
(d) Dependence on $\mathcal{D}_{12}$ of the time-averaged lengths
$\langle L_\phi \rangle$ (red) and $\langle L_\psi \rangle$ (blue), together with the integral-scale Reynolds number $Re$ (purple curve and right vertical axis). The magnitude $J(t)$ of the nonreciprocal flux~\eqref{eq:fluxJ},
for (e) $\mathcal{D}_{12} = 0.12, 0.16, 0.24,$ and $0.33$, with fixed $\nu=0.0001$, and (f) for $\nu = 12, 0.01, 0.001,$ and $0.0001$, with fixed $\mathcal{D}_{12}=0.33$. }
 \label{fig:energy_time}
\end{figure*}
%%%%%%%%%%%%%%%%%%%%%%%%%%%%%%%%%%

\section {Visualisation of Fields}
\label{sec:fields}
To obtain a qualitative characterisation of NRCHNS turbulence, it is instructive to begin with flow visualisation, via pseudocolor plots of $\omega$, and associated plots of the CH fields $\phi$ and $\psi$. In Fig.~\ref{fig:pcolor} we give such pseudocolor plots, at representative times for $\nu=0.0001$, $\sigma_1=\sigma_2=3$, and the four representative values of the non-reciprocity parameter $\mathcal{D}_{12}=0.12$ (column 1), $\mathcal{D}_{12}=0.16$ (column 2), $\mathcal{D}_{12}=0.24$ (column 3), and $\mathcal{D}_{12}=0.33$ (column 4). These plots, for $\omega$ [Figs.~\ref{fig:pcolor} (a)-(d)], $\phi$ [Figs.~\ref{fig:pcolor}(e)-(h)], and $\psi$ [Figs.~\ref{fig:pcolor} (i)-(l)], illustrate how 2D NRCHNS turbulence develops as the strength of the non-reciprocity $\mathcal{D}_{12}$ increases; this is the source of activity in our system. The full spatiotemporal evolution of these fields can be appreciated by looking at the Videos V1-V4 [see the Supplementary Material].

Figure~\ref{fig:subcolor} shows that, with all other parameters held fixed, NRCHNS turbulence is enhanced if we decrease $\nu$, which increases $Re$ [see Refs.~\cite{boffetta2012two,pandit2017overview} for conventional 2D fluid turbulence]. We show, at a representative time and for $\mathcal{D}_{12}=0.33$ and $\sigma_1=\sigma_2=3$, that, as we decrease $\nu$ [$\nu=12, 0.01, 0.001$ and $0.0001$ in columns 1, 2, 3, and 4, respectively], the pseudocolor plots of $\omega$ [Figs.~\ref{fig:subcolor} (a)-(d)], $\phi$ [Figs.~\ref{fig:subcolor} (e)-(h)], and $\psi$ [Figs.~\ref{fig:subcolor} (i)-(l)] display increasing signatures of chaotic fluid motion with vorticity. We note, \textit{en passant}, that waves appear at low $Re$ [see Figs.~\ref{fig:subcolor} (a), (e), and (i) for $\nu = 12$], 
as noted in low-$Re$ studies~\cite{you2020nonreciprocity,saha2020scalar,frohoff2023nonreciprocal,brauns2024nonreciprocal}. A full appreciation of spatiotemporal chaos and turbulence in our 2D NRCHNS system can be obtained by viewing the Videos V5-V7, and V4 [see the Supplementary Material]. 
\section {Temporal Fluctuations}
\label{sec:temporal}

To go beyond the qualitative visualisation of NRCHNS turbulence given in Figs.~\ref{fig:pcolor} and \ref{fig:subcolor}, we use statistical measures that are employed in conventional 2D fluid turbulence~\cite{padhan2025cahn,boffetta2012two,pandit2017overview}. 

Consider first the evolution with the scaled time $t\Omega_L$ of the fluid energy $\mathfrak{E}(t)$ [Fig.~\ref{fig:energy_time} (a)] and the lengths $L_\phi(t)$ [Fig.~\ref{fig:energy_time} (b)] and $L_\psi(t)$ [Fig.~\ref{fig:energy_time} (c)], which are the scales, in the $\phi$ and $\psi$ fields, respectively, at which phase separation is arrested by turbulence [see Eq.~\eqref{eq:specetc} and, for 2D fluid turbulence, Refs.~\cite{perlekar2017two,padhan2025cahn}]. These plots show that $\mathfrak{E}(t)$, $L_\phi(t)$, and $L_\psi(t)$ display fluctuations, because of NRCHNS turbulence; but they approach well-defined mean values at large times, when our system settles into a non-equilibrium statistically steady state (NESS). The time averages $\langle L_\phi(t) \rangle_t$ and $\langle L_\psi(t) \rangle_t$, in this NESS, decrease as we increase the scaled non-reciprocity parameter $\mathcal{D}_{12}$ and, thereby, increase the Reynolds number $Re$ [see Fig.~\ref{fig:energy_time} (d)]. Thus, as NRCHNS turbulence increases, it leads to the suppression of phase separation (often referred to as \textit{coarsening arrest}), in much the same way as do 2D fluid turbulence~\cite{perlekar2017two,padhan2025cahn} and different types of active-fluid turbulence~\cite{padhan2024novel,padhan2025cahn,maji2025emergent}.
There is, however, one crucial difference between NRCHNS turbulence and the other types of 2D turbulence that we have mentioned above. This is best uncovered by computing the magnitude of the non-reciprocal flux, namely, $J(t)$ [Eq.~\eqref{eq:fluxJ}] that has, so far, been used~\cite{rana2024defect} only for low-$Re$ flows. As we increase $Re$ by enhancing $\mathcal{D}_{12}$ [see Fig.~\ref{fig:energy_time}(d)], $J(t)$  begins to display rapid fluctuations as shown in Fig.~\ref{fig:energy_time}(e).  Figure~\ref{fig:energy_time}(f) illustrates the dependence of $J(t)$ on the kinematic viscosity $\nu$. If this is large (say $\nu=12$), the system exhibits propagating waves [Fig.~\ref{fig:subcolor} (a), (e), and (i)] and $J(t)$ attains a finite, steady value [green curve in Fig.~\ref{fig:energy_time}(f)]. As $\nu$ decreases, so does $J(t)$, because $Re$ increases [red, orange, and blue curves in Fig.~\ref{fig:energy_time}(f)]. We emphasize that non-reciprocality induces turbulence; but this also leads to fluctuations that, in turn, suppress $J(t)$, the principal emergent signature of NRCHNS turbulence. 
This suppression is also associated with coarsening arrest, which is characterised by the decrease of $\langle L_\phi\rangle_t$ and $\langle L_\psi\rangle_t$ with increasing $Re$ [cf. Fig.~\ref{fig:energy_time} (d)]; furthermore, this is reminiscent of coarsening arrest by conventional 2D fluid turbulence in binary-fluid systems~\cite{perlekar2017two,padhan2025cahn}.
%%%%%%%%%%%%%%%%%%%%%%%%%%%%%%%%%
\begin{figure*}[htp]
\centering
\includegraphics[width=1.0\textwidth]{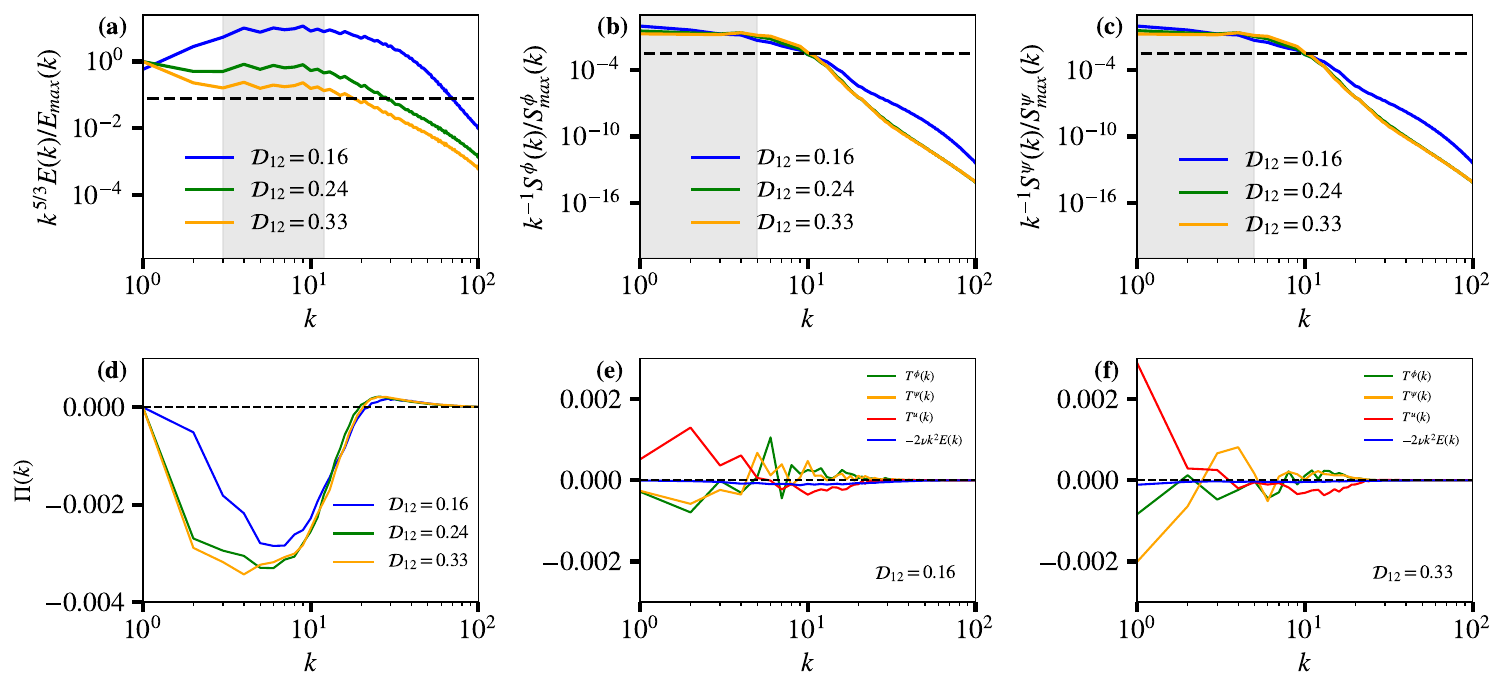}
    \caption{\textbf{The dependence on the wavenumber $k$ of energy and concentration spectra, spectral transfers, and fluxes for $\nu = 0.0001$:} {Log-log plots of compensated spectra versus the scaled wavenumber  $k$, for $\mathcal{D}_{12}=0.16,\,0.24,\,0.33$, with power-law scaling regions shaded in gray: (a) $k^{5/3}E(k)$ [scaling region $3 \lesssim k \lesssim 13$]; (b) $k^{-1}S^{\phi}(k)$ [scaling region $1 \lesssim k \lesssim 4$]; (c) $k^{-1}S^{\psi}(k)$ [scaling region $1 \lesssim k \lesssim 4$].
    %for the spatiotemporal evolution of these pseudocolor plots, see Supplementary Movies V2-V4. 
     For plots of the energy  flux, $\Pi(k)$ , see Figs.(d) for different $\mathcal{D}_{12}=0.16, 0.24$, and $0.33$}. Log-lin plots of $T^u(k)$(red), $T^{\phi}(k)$(green), $ T^{\psi}(k)$(orange), and $2\nu k^2 E(k)$(blue)  [see Eq.~\eqref{eq:spectralbalance}] versus $ k$, in the NESS, for $\alpha=0.0$ and (e) $\mathcal{D}_{12}=0.16$ and (f) $\mathcal{D}_{12}=0.33$.}
     \label{fig:energy_flux}
\end{figure*}
%%%%%%%%%%%%%%%%%%%%%%%%%%%%%%%%%%
\section {Spectra, Spectral Transfers, and Fluxes}
\label{sec:spectral}
The fluid energy, $\phi$, and $\psi$ spectra, $E(k)$, $S^\phi(k)$, and $S^\psi(k)$, respectively, are related to spatial Fourier transforms of the fields $\bm{u}$, $\phi$, and $\psi$ [Eq.~\eqref{eq:specetc}]. In turbulent fluids, fluctuations of the velocity are spread over a wide range of spatial scales. In particular, the dependencies of these spectra on the wave number $k$ help us to identify power-law scaling regions that are often found in different turbulent fluids [see, e.g., Refs.~\cite{boffetta2012two,pandit2017overview,padhan2025cahn}].
If one of these spectra, say $E(k)$, were to display a power-law scaling region $E(k) \sim k^{-\beta}$, then a log-log plot of the \textit{compensated spectrum} $k^\beta E(k)$ would show a flat region in this scaling range.

Does the non-reciprocal turbulence in our NRCHNS model exhibit such power-law scaling? It does, indeed, as we show in Figs.~\ref{fig:energy_flux} (a), (b), and (c), which are log–log plots versus $k$ of the compensated 
spectra $k^{5/3}E(k)$, $k^{-1}S^{\phi}(k)$, and $k^{-1}S^{\psi}(k)$, respectively, for $\mathcal{D}_{12}=0.16, 0.24,$ and $0.33$ and $\nu = 0.0001$. In the shaded gray regions, at low and intermediate $k$, these spectra exhibit extended plateaux, which are indicated by dashed lines. Thus, in these limited scaling regimes, our data are consistent with the power-law forms $E(k)\sim k^{-5/3}$, $S^{\phi}(k)\sim k$, and $S^{\psi}(k)\sim k$. The scaling of $E(k)$ is reminiscent of the inverse-cascade regime in 2D fluid turbulence~\cite{boffetta2012two,pandit2017overview}. 

To explore this similarity, we turn to the spectral energy flux $\Pi(k)$ and spectral balances [Eqs.~\eqref{eq:specetc}-\eqref{eq:spectralbalance}]. Figure~\ref{fig:energy_flux}(d) shows $\Pi(k)$ as a function of $k$ for $\mathcal{D}_{12}=0.16$, $0.24$, and $0.33$ and $\nu=0.0001$; over the same range of $k$ in which $E(k)\sim k^{-5/3}$ [cf. Figs.~\ref{fig:energy_flux} (a) and (d)], we see that $\Pi(k)$ remains approximately constant and negative, indicating that 2D NRCHNS turbulence has an inverse-cascade range [cf. Refs.~\cite{boffetta2012two,pandit2017overview,padhan2024novel}].
%%%%%%%%%%%%%%%%%%%%%%%%%%%%%%%%
%%%%%%%%%%%%%%%%%%%%%%%%%%%%%%%%%%%%
\begin{figure*}[htp]
    \centering
\includegraphics[width=1.0\linewidth]{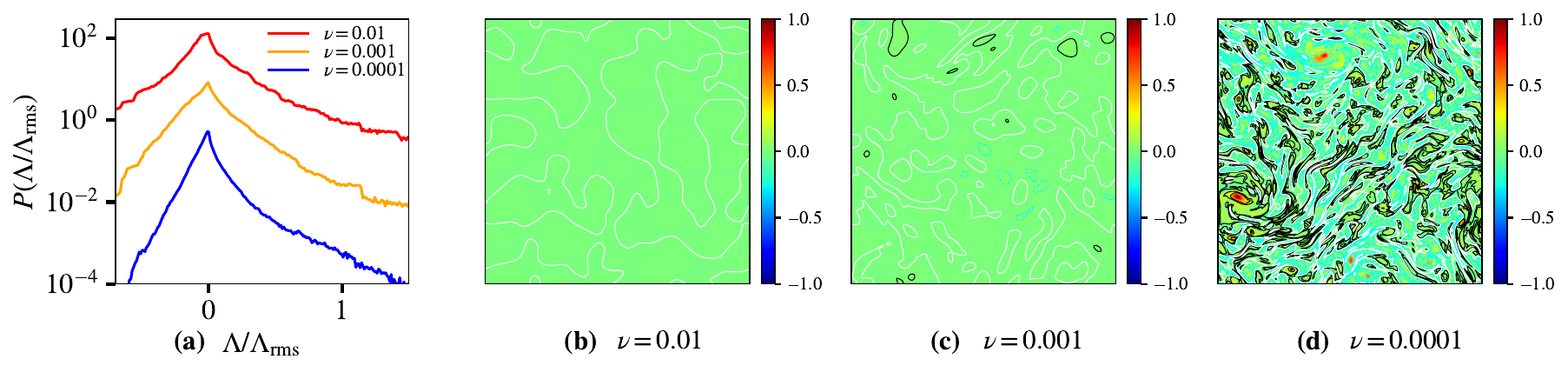}
    \caption{\textcolor{black}{\textbf{Flow topology in NRCHNS turbulence for $\mathcal{D}_{12}=0.33$:} (a) Normalised probability distribution functions (PDFs) of the Okubo–Weiss parameter $\Lambda$ [see Eq.~\eqref{eq:Okubo}] for $\nu=0.01, 0.001$, and $0.0001$ (the top two curves have been translated upwards 
    for clarity). Pseudocolor plots of the normalised Okubo-Weiss paramemter $\Lambda/\Lambda_{max}$ for (b) $\nu=0.01$, (c) $\nu=0.001$, and (d) $\nu=0.0001$; contours of the normalized vorticity $\omega/\omega_{max}=-0.6$ (dashed-brown contours), $\omega/\omega_{max}=0$ (white contours), and $\omega/\omega_{max}=0.6$ (black contours) are superimposed.}}
    \label{fig:pdf_okb}
\end{figure*}
%%%%%%%%%%%%%%%%%%%%%%%%%%%%%%%%%
We now investigate how energy is transferred across scales in the nonequilibrium steady state (NESS) of 2D NRCHNS turbulence. Our analysis is based on the spectral energy-balance relation~\eqref{eq:spectralbalance}, which allows us to quantify the individual contributions to the energy budget at each wavenumber $k$. In Figs.~\ref{fig:energy_flux}(e) and (f), we display 
the nonlinear inertial transfer $T^{u}(k)$ (red), viscous dissipation $2\nu k^{2}E(k)$ (blue), and the stress-induced transfers associated with the $\phi$ and $\psi$ fields, denoted by $T^{\phi}(k)$ (green) and $T^{\psi}(k)$ (orange), respectively, for the two representative values $\mathcal{D}_{12}=0.16$ and $0.33$; here, $\nu = 0.0001$. 
From these plots we see that the dominant spectral balance in the relation~\eqref{eq:spectralbalance} occurs between the terms  $T^{u}(k)$, $T^{\phi}(k)$, and $T^{\psi}(k)$; the viscous contribution $2\nu k^2 E(k)$ is much smaller than these terms in the range of $k$ that we present. 
\section {Flow Topology}
\label{sec:topology}
Our DNS also provides important insights into the topology of NRCHNS turbulent flows. We characterize this topology using the Okubo–Weiss parameter $\Lambda$, which has been used since the pioneering suggestions of Okubo~\cite{okubo1970horizontal} and Weiss~\cite{weiss1991dynamics}, in a variety of 2D fluid-dynamical systems [see, e.g., Refs.~\cite{perlekar2009statistically,pandit2017overview,kiran2023irreversibility,mukherjee2023intermittency,padhan2025interfaces,maji2025emergent}]: 
\begin{equation}
\Lambda(x,y) \equiv \frac{\bm{\omega}^2(x,y) - \bm{\Sigma}^2(x,y)}{8}\,,
\label{eq:Okubo}
\end{equation}
where $\bm{\Sigma}$ is the symmetric part of the velocity-derivative tensor $\partial_iu_j$; clearly, $\Lambda > 0$ in the vortical regions of the flow and $\Lambda < 0$ in extensional parts.
We compute $\Lambda$ from our DNS of the NRCHNS system~\eqref{eq:phi}-\eqref{eq:incom} and then obtain its probability distribution function (PDF) $\mathcal{P}(\Lambda)$. Figure~\ref{fig:pdf_okb}(a) shows representative plots of  $\mathcal{P}(\Lambda)$ for $\mathcal{D}_{12}=0.33$ and $\nu=0.01,\, 0.001$, and $0.0001$. From the definition of $\Lambda$, we see that its mean  $\langle \Lambda \rangle = 0$. In the nonequilibrium steady state of NRCHNS turbulence ($\nu \lesssim 0.01 $), the PDF $\mathcal{P}(\Lambda)$ develops a cusp at $\Lambda=0$ and shows significant skewness, which is similar to that observed in 2D fluid turbulence~\cite{perlekar2009statistically,pandit2017overview} and in the 2D Toner–Tu–Swift–Hohenberg (TTSH) model for bacterial turbulence~\cite{kiran2023irreversibility,mukherjee2023intermittency}, and in other types of CHNS turbulence~\cite{padhan2025interfaces,maji2025emergent}. This asymmetry in $\mathcal{P}(\Lambda)$ grows significantly as $\nu$ is reduced from $0.001$ to $0.0001$, i.e., as $Re$ increases, the dominance of vortical over extensional regions is enhanced.

It is instructive to superimpose contours of the vorticity $\omega$ on pseudocolor plots of $\Lambda$. We present such superposed plots in Figs.~\ref{fig:pdf_okb} (b)–(d) for $\mathcal{D}_{12}=0.33$ and $\nu=0.01,\, 0.001$, and $0.0001$, with contours of the normalised vorticity $\omega/\omega_{max}=-0.6$ (brown dashed), $\omega/\omega_{max}=0$ (white), and $\omega/\omega_{max}=0.6$ (black). These visualisations demonstrate how $\Lambda$ segregates the flow into rotation-dominated ($\Lambda>0$) and strain-dominated ($\Lambda<0$) regions. The resulting flow topology resembles that observed in 2D fluid turbulence~\cite{perlekar2009statistically,daniel2002topology} and other types of CHNS turbulence~\cite{padhan2025interfaces,maji2025emergent}.
%%%%%%%%%%%%%%%%%%%%%%%%%%%%%%%
\begin{figure}[htp]
    \centering
   \includegraphics[width=1.0\linewidth]{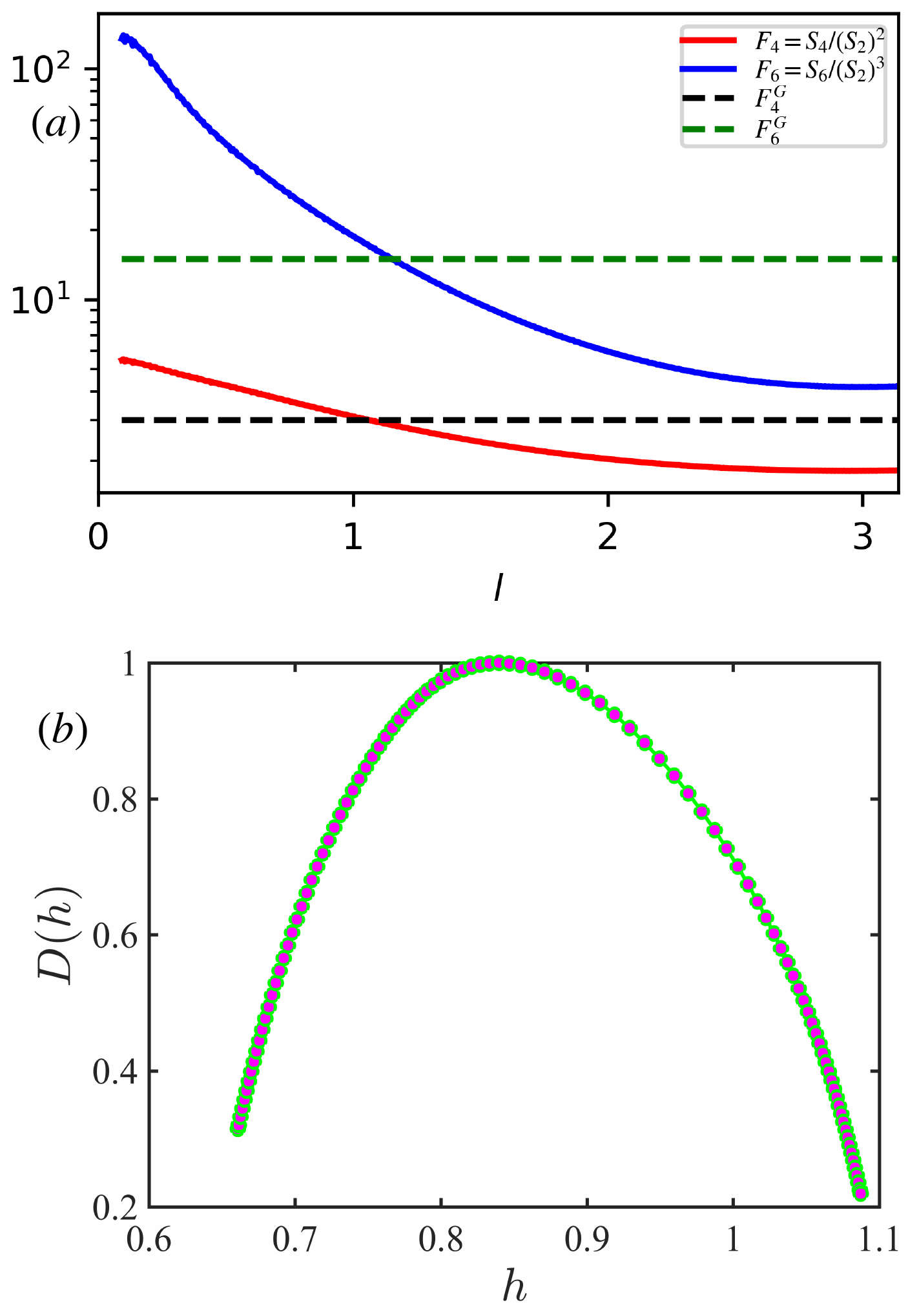}
    \caption{\textcolor{black}{\textbf{Plots of the flatness, hyper-flatness, and multifractal spectrum for NRCHNS turbulence with $\mathcal{D}_{12}=0.33$:}
    (a) Semilog plots the flatness $F_4 \equiv S_4/(S_2)^2$ and the hyper-flatness $F_6 \equiv S_6/(S_2)^3$ versus the length scale $l$ reveal clear departures from their Gaussian values ($3$ and $15$, indicated by black and green dashed lines, respectively) at small length scales $l$. (b) The multifractal spectrum $D(h)$ as a function of the Hurst exponent $h$, obtained from $J(t)$ (see text).}}
    \label{fig:str}
\end{figure}
%%%%%%%%%%%%%%%%%%%%%%%%%%%%%%%%%%%
\section{Multifractality of NRCHNS Turbulence}
\label{sec:mult}

Multifractality is a defining feature of fluid turbulence [see, e.g., Refs.~\cite{frisch1995turbulence,meneveau1991multifractal,boffetta2008twenty}]. It is natural, therefore, to ask whether NRCHNS turbulence also shows signatures of multifractality. To address this question, we begin with the longitudinal velocity increments $\delta u_\parallel(l)$, the order-$p$ structure function $\mathfrak{S}_p$, the flatness $F_4$, and the hyperflatness $F_6(l)$, which are defined as follows:
\begin{eqnarray}
\delta u_\parallel(\bm{r},l) &\equiv& [\bm{u}(\bm{r}+\bm{l})-\bm{u}(\bm{r})]\cdot \bm{l}/|\bm{l}|\,; \nonumber \\
\mathfrak{S}_p(l) &\equiv& \langle [\delta u_\parallel(\bm{r},l)]^p\rangle_{\bm{r}}\,; \nonumber \\
F_4(l) &\equiv& \mathfrak{S}_4(l)/[\mathfrak{S}_2(l)]^2 \,; \nonumber \\
F_6(l) &\equiv& \mathfrak{S}_6(l)/[\mathfrak{S}_2(l)]^3\,.
\end{eqnarray}
In Fig.~\ref{fig:str} (a) we plot the flatness $F_4(l)$ and hyperflatness $F_6(l)$ versus $l$; we see that both
$F_4(l)$ and $F_6(l)$ rise well above their Gaussian values of $3$ and $15$, respectively, at small length scales $l$; this is a clear signature of small-scale intermittency. Furthermore, we show that the fluctuations in $J(t)$ [see Fig.~\ref{fig:energy_time} (e)] exhibit temporal multifractality, in so far as they yield a broad multifractal spectrum $D(h)$ [see Fig.~\ref{fig:str} (b)], which we have evaluated using the dwtleader package in Matlab [cf. Ref.~\cite{padhan2025cahn}]. 

\section{Discussion}
\label{sec:disc}
 Non-reciprocal effective interactions can lead to remarkable effects that continue to be explored in fields as diverse as soft and active matter, and neuronal, chemical, biological, and non-Hermitian systems, To the best of our knowledge, non-reciprocal hydrodynamical turbulence has, so far, been a \textit{terra incognita}. Therefore, we have initiated an exploration of such turbulence. We have shown that non-reciprocal binary-fluid systems provide ideal candidates for the exploration of such turbulence. In particular, we have developed the 2D NRCHNS model~\eqref{eq:phi}-\eqref{eq:incom} and demonstrated, via extensive DNSs, that, as we increase $\mathcal{D}_{12}$, the strength of the non-reciprocity, we have obtained a new type of turbulence, with an inverse-cascade energy spectrum $E(k)\sim k^{-5/3}$, an energy flux $\Pi(k)$ that is nearly constant and negative for $k$ in this inverse-cascade range. Although inverse cascades occur in forced, 2D fluid turbulence~\cite{boffetta2012two,pandit2017overview} and other active-fluid systems~\cite{kiran2023irreversibility,mukherjee2023intermittency,padhan2024novel}, they are distinct from NRCHNS turbulence, which is characterised by the non-reciprocal flux with magnitude $J(t) > 0$. 
 
Experimentally non-reciprocal chaos has been observed in different systems; recent examples include a system with physical agents that have dynamic binary ideologies~\cite{li2024memory} and a non-reciprocal active-droplet system~\cite{liu2024self}. The latter example might well be best suited for experimental studies of the type of non-reciprocal turbulence that we have elucidated.

A recent preprint~\cite{klamser2025directed} has presented a numerical study of a directed-percolation transition, in dense particle assemblies with non-reciprocal pairwise forces, to a state with active turbulence. It will be interesting to compare this turbulence with the non-reciprocal turbulence that we have found in our hydrodynamical NRCHNS model.

\section{Model and Methods}
\label{sec:ModMeth}
Our minimal hydrodynamical model, the NRCHNS system, builds on the model of Ref.~\cite{padhan2023activity,sabrina2015coarsening} and combines it with the non-reciprocal Cahn-Hilliard models of Refs.~\cite{saha2020scalar,frohoff2023non,frohoff2023nonreciprocal,brauns2024nonreciprocal}. The NRCHNS system 
that we use is governed by following the set of partial differential equations (PDEs), which we write in terms of the vorticity $\bm{\omega} = (\nabla \times \bm u) = \omega \hat{z}$. 
\begin{eqnarray}
\mathcal{F}
\equiv \int_\Omega[ \frac{3}{16} \{\frac{\sigma_1}{\epsilon_1}(\phi^2 - 1)^2 &+& \frac{\sigma_2}{\epsilon_2}(\psi^2 - 1)^2\} \nonumber \\
+\frac{3}{4}\{\sigma_1 \epsilon_1 |\nabla \phi|^2 &+& \sigma_2 \epsilon_2 |\nabla \psi|^2\}] d\Omega\,; \label{eq:functional} \\
    \partial_t \phi + (\bm u \cdot \nabla) \phi = M_1 \nabla^2 \left( \frac{\delta \mathcal F}{\delta \phi}\right) &+& M_{1}\nabla^2 D_{12}\psi\,;\label{eq:phiA} \\
    \partial_t \psi + (\bm u \cdot \nabla) \psi = M_2 \nabla^2 \left( \frac{\delta \mathcal F}{\delta \psi}\right) &+& M_{2}\nabla^2 D_{21}\phi\,;\label{eq:psiA} \\
    \partial_t \omega + (\bm u \cdot \nabla) \omega &=& [\nu \nabla^2 \omega  \nonumber \\
    -\frac{3}{2}\sigma_1\epsilon_1\nabla\times\mathcal (\nabla^2\phi\nabla\phi)&-& \frac{3}{2}\sigma_2\epsilon_2\nabla\times\mathcal (\nabla^2\psi\nabla\psi) \nonumber \\
    &-&\alpha\omega]\,;\label{eq:omegaA} \\
    \nabla \cdot \bm u &=& 0\,;\label{eq:incomA}
     \end{eqnarray}
Note that the terms with the coefficients $D_{12}$ and $D_{21}$, which are responsible for the non-reciprocal nature of this model, do not follow from a functional derivative of the free-energy functional $\mathcal{F}$. The above equations can be written in the equivalent form given in Eqs.~\eqref{eq:phi}-~\eqref{eq:incom}.

% Clearly, the non-reciprocal terms, with the coefficients $D_{12}$ and $D_{21}$, do not follow from functional derivatives of $\mathcal{F}$.

We solve Eqs.~\eqref{eq:phi}-\eqref{eq:incom}, in a square domain, with periodic boundary conditions in both spatial directions, using a Fourier pseudospectral direct numerical simulation (DNS) that has been described in Refs.~\cite{padhan2024novel,padhan2025cahn,maji2025emergent}. In such a DNS, spatial derivatives are evaluated simply in Fourier space and products in physical space; we employ the $1/2$ rule for the suppression of aliasing errors; our domain contains $N^2$ collocation points.  For time integration, we employ the semi-implicit ETDRK-2 method~\cite{cox2002exponential}. We have developed a CUDA C code on GPU processors. The parameter values that we use in our DNS are as follows: $N=1024$, $M_1=M_2=0.0001$, $\epsilon_1=\epsilon_2=0.01839$, $\sigma_1=\sigma_2=3$,  and $\alpha=0.0$. We have also carried out several runs with $N=512$, to check that our results so not depend significantly on the values of $N$ we use. We list non-dimensional parameters in Table~\ref{tab:param}.

We begin with a state in which both $\phi$ and $\psi$ have zero mean; specifically, we choose $\phi(x,y,t=0)$ and $\psi(x,y,t=0)$ to be independent and identically distributed random numbers, at every point $(x,y)$, that are distributed uniformly on the interval $[-0.1, 0.1]$; we start with zero vorticity, i.e., $\omega(x,y,t=0) = 0$. 

From our DNS we obtain the spatiotemporal evolution of the fields $\bm{u}$, $\phi$, and $\psi$; their spatial Fourier transforms (denoted by tildes) yield the following fluid-energy $[\mathcal{E}(k,t), E(k)]$ and concentration $[\mathcal{S}^\phi(k,t), S^\phi(k), \mathcal{S}^\psi(kt), S^\psi(k)]$ spectra, the fluid energy [$\mathfrak{E}(t)$], the turbulent integral length scale $L$, and the coarsening-arrest length scales $L_\phi$ and $L_\psi$:
\begin{eqnarray}
   \mathcal{E}(k,t)&\equiv&\frac{1}{2}\sum_{k'= k-1/2}^{k'= k+ 1/2}|\tilde{\bm{u}}(\textbf{k}',t)|^2; \; E(k) \equiv \langle \mathcal{E}(k,t) \rangle_t\,;\label{eq:Ek} \nonumber \\
   \mathfrak{E}(t) &\equiv&\sum_k \mathcal{E}(k,t)\,;\quad L\equiv \frac{\sum_k k^{-1}E(k)}{\sum_k E(k)}\,; \label{eq:EtL} \nonumber \\
   \mathcal{S}^\phi(k,t)&\equiv&\frac{1}{2}\sum_{k'= k-1/2}^{k'= k+ 1/2}|\tilde{\phi}(\textbf{k}',t)|^2; \; S^\phi(k) \equiv \langle \mathcal{S}^\phi(k,t) \rangle_t\,; \label{eq:Skphi} \nonumber \\
   \mathcal{S}^\psi(k,t)&\equiv&\frac{1}{2}\sum_{k'= k-1/2}^{k'= k+ 1/2}|\tilde{\psi}(\textbf{k}',t)|^2; \; \mathcal{S}^\psi(k) \equiv \langle S^\psi(k,t) \rangle_t\,; \label{eq:Skpsi} \nonumber \\
   L_{\phi} &\equiv& \frac{\sum_k S^\phi(k)}{\sum_k k S^\phi(k)}\,; \quad L_{\psi} \equiv \frac{\sum_k S^\psi(k)}{\sum_k k S^\psi(k)}\,; \nonumber \\
   \label{eq:specetc}
\end{eqnarray}
    the wave vectors $\mathbf{k}$ and $\mathbf{k}'$ have moduli $k$ and $k'$, respectively; $\langle\cdot\rangle_t$ denotes the time average. We use the integral-scale frequency $\Omega_L \equiv u_{rms}/L$ to non-dimensionalize time; the root-mean-square velocity $u_{rms}\equiv \sqrt{2\langle \mathfrak{E}(t) \rangle_t}$. The emergent  non-reciprocity can be characterized~\cite{rana2024defect} by the magnitude $J(t)=\sqrt{J_x^2(t)+J_y^2(t)}$ of the net flux~\eqref{eq:fluxJ}.
    
\begin{table}[h]
    \centering
\renewcommand{\arraystretch}{1.3} %Adjust row height
\setlength{\tabcolsep}{6pt} % Adjust column spacing
\begin{tabular}{|c|c|c|c|c|c|c|} 
 \hline
  $\mathcal{D}_{12}$ & $\nu$  & $We$ & $Cn$ & $Pe$ & $Re$  \\ 
 \hline
 0.12 & $10^{-4}$ & 0.0289 & 0.067 & 9.50 & 1549.5 \\ 
  \hline
 0.16 &  $10^{-4}$    &  0.0330 & 0.061 & 10.53 & 1717.3 \\ 
 \hline
 0.24 & $10^{-4}$    & 0.1480 & 0.024& 35.50  &  5786.7 \\ 
 \hline
 0.33 & $10^{-4}$    & 0.3948 & 0.020 & 63.45  & 10341.2 \\ 
 \hline
 0.33 &  $12$   & $\mathcal{O}(10^{-14})$ & 0.074 & $\mathcal{O}(10^{-5})$ &  $\mathcal{O}(10^{-9})$ \\
 \hline
 0.33 & $10^{-2}$   & 0.0010 & 0.036 & 2.48 &  4.04 \\ 
 \hline
 0.33 & $10^{-3}$   & 0.0624 & 0.021 & 25.39 &  413.81 \\ 
 \hline
\end{tabular}
\caption{Table of the values of the non-dimensional parameters, for different values of $\mathcal{D}_{12}$ [column 1]: Cahn number $Cn_1 \equiv \epsilon_1/L $, $Cn_2 \equiv \epsilon_2/L $, with $Cn=Cn_1=Cn_2$, Weber number $We_1 \equiv L u_{rms}^2/\sigma_1$, $We_2 \equiv L u_{rms}^2/\sigma_2$, with $We=We_1=We_2$, P\'eclet number $ Pe_1 \equiv Lu_{rms} \epsilon_1/M_1\sigma_1$, $ Pe_2 \equiv Lu_{rms} \epsilon_2/M_2\sigma_2$, with $Pe=Pe_1=Pe_2$  and friction $\alpha'=\alpha L /u_{rms}$, where $u_{rms}$ is the root-mean-square velocity. We use $\alpha = 0$.}
 \label{tab:param}
\end{table}
In addition to $\mathcal{D}_{12}\equiv |D_{12}|/D_{11}$, the following control parameters [see Table~\ref{tab:param}] govern the nonequilibrium states of our 2D NRCHNS system: the integral-scale Reynolds $Re \equiv L u_{rms}/{\nu}$, Cahn $Cn \equiv \epsilon/L $, Weber $We \equiv L u_{rms}^2/\sigma$, and P\'eclet $ Pe \equiv Lu_{rms} \epsilon/M\sigma$ numbers, and the non-dimensionalized friction $\alpha'=\alpha L /u_{rms}$. 
\textcolor{black}{\begin{eqnarray}
    T^u(k)&=&\sum_{k'= k-1/2}^{k'= k+ 1/2} \langle\widetilde{\bm{u}(-\bm{k}'}).\bm{P}(\bm{k}').\widetilde{(\bm{u}.    \grad\bm{u})}(\bm{k}')\rangle_t \,;\nonumber  \\
    T^{\phi}(k)&=&-\sum_{k'= k-1/2}^{k'= k+ 1/2} \langle\widetilde{\bm{u}(-\bm{k}'}).\bm{P}(\bm{k}').\widetilde{(\nabla^2\phi\nabla\phi)}(\bm{k}')\rangle_t \,;\nonumber\\
    T^{\psi}(k)&=&-\sum_{k'= k-1/2}^{k'= k+ 1/2} \langle\widetilde{\bm{u}(-\bm{k}'}).\bm{P}(\bm{k}').\widetilde{(\nabla^2\psi\nabla\psi)}(\bm{k}')\rangle_t \,;\nonumber\\
    \Pi(k)&=&\sum_{k'=0}^{k'=k} T^u(k')\,; 
    \label{eq:Tketc}
\end{eqnarray}}
here, the transverse projector $\bm{P}(\bm{k})$ has the components $P_{ij}(\bm{k})=[\delta_{ij} -(k_ik_j/k^2)]$.
The spectral energy balance is given by ~\cite{padhan2024novel, boffetta2012two,verma2019energy}
\begin{eqnarray}
\partial_t{{\mathcal{E}(k,t)}} &=& - T^u(k,t) + T^{\phi}(k,t) - 2\nu k^2 \mathcal{E}(k,t) \nonumber \\  &+& T^{\psi}(k,t) \,. \label{eq:spectralbalance}
\end{eqnarray}

\section*{Acknowledgments}
   BM and RP thank A. Basu and K.V. Kiran for discussions, the Anusandhan National Research Foundation (ANRF), the Science and Engineering Research Board (SERB), and the National Supercomputing Mission (NSM), India, for support,  and the Supercomputer Education and Research Centre (IISc), for computational resources. BM, NBP, and RP would like to thank the Isaac Newton Institute for Mathematical Sciences, Cambridge, for support and hospitality during the programme \textit{Anti-diffusive dynamics\,: from sub-cellular to astrophysical scales (EPSRC grant no. EP/R014604/1)}, where some of the work on this paper was undertaken. NBP and AV acknowledge financial support from the German Research Foundation (DFG) through the project "Analysing structure-property relations in equilibrium and non-equilibrium hyperuniform systems" (project number VO 899/32-1).
\section*{Appendix}
\subsection*{Videos}
The following videos show the spatiotemporal evolution of the pseudocolor plots in Figs. ~\ref{fig:pcolor}, ~\ref{fig:subcolor} 
\begin{itemize}
\item Video V1: This shows the spatiotemporal evolution of pseudocolor renderings [cf. Figs. 1 (a), (e), (i)] of the fields  $\omega$ , $\phi$ and $\psi$ for $\mathcal{D}_{12}=0.12$ and $\nu=0.0001$ \textcolor{blue}{\url{https://youtu.be/TYALU2TTXAY}}; 
\item Video V2: This shows the spatiotemporal evolution of pseudocolor renderings [cf. Figs. 1 (b), (f), (j)] of the fields  $\omega$ , $\phi$ and $\psi$ for $\mathcal{D}_{12}=0.16$ and $\nu=0.0001$ \textcolor{blue}{\url{https://youtu.be/hgFcZWOaycM}}; 
\item Video V3: This shows the spatiotemporal evolution of pseudocolor renderings [cf. Figs. 1 (c), (g), (k)] of the fields  $\omega$ , $\phi$ and $\psi$ for $\mathcal{D}_{12}=0.24$ and $\nu=0.0001$\\ \textcolor{blue}{\url{https://youtu.be/wR639pdXOUs}}; 
\item Video V4: This shows the spatiotemporal evolution of pseudocolor renderings [cf. Figs. 1 (d), (h), (l) and 2 (d), (h), (l)] of the fields  $\omega$ , $\phi$ and $\psi$ for $\mathcal{D}_{12}=0.33$ and $\nu=0.0001$
\\ \textcolor{blue}{\url{https://youtu.be/3upOUEgZNm4}}; 
\item Video V5: This shows the spatiotemporal evolution of pseudocolor renderings [cf. Figs. 2 (a), (e), (i)] of the fields  $\omega$ , $\phi$ and $\psi$ for $\mathcal{D}_{12}=0.33$ and $\nu=12$ \textcolor{blue}{\url{https://youtu.be/9kfpGyYCHZE}};
\item Video V6: This shows the spatiotemporal evolution of pseudocolor renderings [cf. Figs. 2 (b), (f), (j)] of the fields  $\omega$ , $\phi$ and $\psi$ for $\mathcal{D}_{12}=0.33$ and $\nu=0.01$ \textcolor{blue}{\url{https://youtu.be/krh7OTY4g5E}};
\item Video V7: This shows the spatiotemporal evolution of pseudocolor renderings [cf. Figs. 2 (c), (g), (k)] of the fields  $\omega$ , $\phi$ and $\psi$ for $\mathcal{D}_{12}=0.33$ and $\nu=0.001$\\ \textcolor{blue}{\url{https://youtu.be/a9cfa52mpL4}};
\end{itemize}
\bibliography{main}

@PREAMBLE{
 "\providecommand{\noopsort}[1]{}" 
 # "\providecommand{\singleletter}[1]{#1}%" 
}

@article{padhan2023activity,
  title={Activity-induced droplet propulsion and multifractality},
  author={Padhan, Nadia Bihari and Pandit, Rahul},
  journal={Physical Review Research},
  volume={5},
  number={3},
  pages={L032013},
  year={2023},
  publisher={APS}
}

@article{padhan2024novel,
  title={Novel turbulence and coarsening arrest in active-scalar fluids},
  author={Padhan, Nadia Bihari and Kiran, Kolluru Venkata and Pandit, Rahul},
  journal={Soft Matter},
  volume={20},
  number={17},
  pages={3620--3627},
  year={2024},
  publisher={Royal Society of Chemistry}
}

@article{boffetta2008twenty,
  title={Twenty-five years of multifractals in fully developed turbulence: a tribute to Giovanni Paladin},
  author={Boffetta, Guido and Mazzino, A and Vulpiani, Angelo},
  journal={Journal of Physics A: Mathematical and Theoretical},
  volume={41},
  number={36},
  pages={363001},
  year={2008},
  publisher={IOP Publishing}
}

@article{padhan2025cahn,
  title={The Cahn--Hilliard--Navier--Stokes framework for multiphase fluid flows: laminar, turbulent and active},
  author={Padhan, Nadia Bihari and Pandit, Rahul},
  journal={Journal of Fluid Mechanics},
  volume={1010},
  pages={P1},
  year={2025},
  publisher={Cambridge University Press}
}

@article{falkovich2006lessons,
  title={Lessons from hydrodynamic turbulence},
  author={Falkovich, Gregory and Sreenivasan, Katepalli R},
  journal={Physics Today},
  volume={59},
  number={4},
  pages={43--49},
  year={2006},
  publisher={AIP Publishing}
}

@article{sreenivasan2025turbulence,
  title={What is the turbulence problem, and when may we regard it as solved?},
  author={Sreenivasan, Katepalli R and Schumacher, J{\"o}rg},
  journal={Annual Review of Condensed Matter Physics},
  volume={16},
  number={1},
  pages={121--143},
  year={2025},
  publisher={Annual Reviews}
}

@article{miri2019exceptional,
  title={Exceptional points in optics and photonics},
  author={Miri, Mohammad-Ali and Alu, Andrea},
  journal={Science},
  volume={363},
  number={6422},
  pages={eaar7709},
  year={2019},
  publisher={American Association for the Advancement of Science}
}

@article{fruchart2021non,
  title={Non-reciprocal phase transitions},
  author={Fruchart, Michel and Hanai, Ryo and Littlewood, Peter B and Vitelli, Vincenzo},
  journal={Nature},
  volume={592},
  number={7854},
  pages={363--369},
  year={2021},
  publisher={Nature Publishing Group UK London}
}

@article{halatek2018self,
  title={Self-organization principles of intracellular pattern formation},
  author={Halatek, Jacob and Brauns, Fridtjof and Frey, Erwin},
  journal={Philosophical Transactions of the Royal Society B: Biological Sciences},
  volume={373},
  number={1747},
  pages={20170107},
  year={2018},
  publisher={The Royal Society}
}

@article{li2024memory,
  title={Memory and Personality Shape Ideological Polarization},
  author={Li, Shengkai and Phan, Trung V and Di Carlo, Luca and Wang, Gao and Do, Van H and Mikhail, Elia and Austin, Robert H and Liu, Liyu},
  journal={arXiv preprint arXiv:2409.06660},
  year={2024}
}

@article{cox2002exponential,
  title={Exponential time differencing for stiff systems},
  author={Cox, Steven M and Matthews, Paul C},
  journal={Journal of Computational Physics},
  volume={176},
  number={2},
  pages={430--455},
  year={2002},
  publisher={Elsevier}
}

@book{puri2009kinetics,
  title={Kinetics of phase transitions},
  author={Puri, Sanjay and Wadhawan, Vinod},
  year={2009},
  publisher={CRC press}
}

@article{maji2025emergent,
  title={Emergent turbulence and coarsening arrest in active-spinner fluids},
  author={Maji, Biswajit and Padhan, Nadia Bihari and Pandit, Rahul},
  journal={Communications Physics},
  year={2025},
  publisher={Nature Publishing Group UK London}
}

@inproceedings{okubo1970horizontal,
  title={Horizontal dispersion of floatable particles in the vicinity of velocity singularities such as convergences},
  author={Okubo, Akira},
  booktitle={Deep sea research and oceanographic abstracts},
  volume={17},
  pages={445--454},
  year={1970},
  organization={Elsevier}
}

@article{kiran2023irreversibility,
  title={Irreversibility in bacterial turbulence: Insights from the mean-bacterial-velocity model},
  author={Kiran, Kolluru Venkata and Gupta, Anupam and Verma, Akhilesh Kumar and Pandit, Rahul},
  journal={Physical Review Fluids},
  volume={8},
  number={2},
  pages={023102},
  year={2023},
  publisher={APS}
}

@article{mukherjee2023intermittency,
  title={Intermittency, fluctuations and maximal chaos in an emergent universal state of active turbulence},
  author={Mukherjee, Siddhartha and Singh, Rahul K and James, Martin and Ray, Samriddhi Sankar},
  journal={Nature Physics},
  volume={19},
  number={6},
  pages={891--897},
  year={2023},
  publisher={Nature Publishing Group UK London}
}

@article{daniel2002topology,
  title={Topology of two-dimensional turbulence},
  author={Daniel, W Brent and Rutgers, Maarten A},
  journal={Physical review letters},
  volume={89},
  number={13},
  pages={134502},
  year={2002},
  publisher={APS}
}

@article{perlekar2009statistically,
  title={Statistically steady turbulence in thin films: direct numerical simulations with Ekman friction},
  author={Perlekar, Prasad and Pandit, Rahul},
  journal={New Journal of Physics},
  volume={11},
  number={7},
  pages={073003},
  year={2009},
  publisher={IOP Publishing}
}

@article{perlekar2017two,
  title={Two-dimensional turbulence in symmetric binary-fluid mixtures: Coarsening arrest by the inverse cascade},
  author={Perlekar, Prasad and Pal, Nairita and Pandit, Rahul},
  journal={Scientific Reports},
  volume={7},
  number={1},
  pages={44589},
  year={2017},
  publisher={Nature Publishing Group UK London}
}

@article{blom2026dynamic,
  title={Dynamic models for two nonreciprocally coupled fields: A microscopic derivation for zero, one, and two conservation laws},
  author={Blom, Kristian and Thiele, Uwe and Godec, Alja{\v{z}}},
  journal={SciPost Physics},
  volume={20},
  number={1},
  pages={005},
  year={2026}
}

@article{padhan2025interfaces,
  title={Interfaces as transport barriers in two-dimensional Cahn--Hilliard--Navier--Stokes turbulence},
  author={Padhan, Nadia Bihari and Pandit, Rahul},
  journal={Journal of Physics: Condensed Matter},
  volume={37},
  number={47},
  pages={475103},
  year={2025},
  publisher={IOP Publishing}
}

@article{weiss1991dynamics,
  title={The dynamics of enstrophy transfer in two-dimensional hydrodynamics},
  author={Weiss, John},
  journal={Physica D: Nonlinear Phenomena},
  volume={48},
  number={2-3},
  pages={273--294},
  year={1991},
  publisher={Elsevier}
}

@book{verma2019energy,
  title={Energy transfers in fluid flows: multiscale and spectral perspectives},
  author={Verma, Mahendra K},
  year={2019},
  publisher={Cambridge University Press}
}

@article{sabrina2015coarsening,
  title={Coarsening dynamics of binary liquids with active rotation},
  author={Sabrina, Syeda and Spellings, Matthew and Glotzer, Sharon C and Bishop, Kyle JM},
  journal={Soft Matter},
  volume={11},
  number={43},
  pages={8409--8416},
  year={2015},
  publisher={Royal Society of Chemistry}
}

@article{brauns2024nonreciprocal,
  title={Nonreciprocal pattern formation of conserved fields},
  author={Brauns, Fridtjof and Marchetti, M Cristina},
  journal={Physical Review X},
  volume={14},
  number={2},
  pages={021014},
  year={2024},
  publisher={APS}
}

@article{boffetta2012two,
  title={Two-dimensional turbulence},
  author={Boffetta, Guido and Ecke, Robert E},
  journal={Annual review of fluid mechanics},
  volume={44},
  number={1},
  pages={427--451},
  year={2012},
  publisher={Annual Reviews}
}

@article{pandit2017overview,
  title={An overview of the statistical properties of two-dimensional turbulence in fluids with particles, conducting fluids, fluids with polymer additives, binary-fluid mixtures, and superfluids},
  author={Pandit, Rahul and Banerjee, Debarghya and Bhatnagar, Akshay and Brachet, Marc and Gupta, Anupam and Mitra, Dhrubaditya and Pal, Nairita and Perlekar, Prasad and Ray, Samriddhi Sankar and Shukla, Vishwanath and others},
  journal={Physics of fluids},
  volume={29},
  number={11},
  year={2017},
  publisher={AIP Publishing}
}

@article{frohoff2023nonreciprocal,
  title={Nonreciprocal Cahn-Hilliard model emerges as a universal amplitude equation},
  author={Frohoff-H{\"u}lsmann, Tobias and Thiele, Uwe},
  journal={Physical review letters},
  volume={131},
  number={10},
  pages={107201},
  year={2023},
  publisher={APS}
}

@article{frohoff2023non,
  title={Non-reciprocity induces resonances in a two-field Cahn--Hilliard model},
  author={Frohoff-H{\"u}lsmann, Tobias and Thiele, Uwe and Pismen, Len M},
  journal={Philosophical Transactions of the Royal Society A},
  volume={381},
  number={2245},
  pages={20220087},
  year={2023},
  publisher={The Royal Society}
}

@article{pisegna2025can,
  title={Can hydrodynamic interactions destroy travelling waves formed by non-reciprocity?},
  author={Pisegna, Giulia and Rana, Navdeep and Golestanian, Ramin and Saha, Suropriya},
  journal={arXiv e-prints},
  pages={arXiv--2501},
  year={2025}
}

@article{gompper20202020,
  title={The 2020 motile active matter roadmap},
  author={Gompper, Gerhard and Winkler, Roland G and Speck, Thomas and Solon, Alexandre and Nardini, Cesare and Peruani, Fernando and L{\"o}wen, Hartmut and Golestanian, Ramin and Kaupp, U Benjamin and Alvarez, Luis and others},
  journal={Journal of Physics: Condensed Matter},
  volume={32},
  number={19},
  pages={193001},
  year={2020},
  publisher={IOP Publishing}
}

@article{tucci2024nonreciprocal,
  title={Nonreciprocal collective dynamics in a mixture of phoretic Janus colloids},
  author={Tucci, Gennaro and Golestanian, Ramin and Saha, Suropriya},
  journal={New Journal of Physics},
  volume={26},
  number={7},
  pages={073006},
  year={2024},
  publisher={IOP Publishing}
}

@article{rana2024defect,
  title={Defect interactions in the non-reciprocal Cahn--Hilliard model},
  author={Rana, Navdeep and Golestanian, Ramin},
  journal={New Journal of Physics},
  volume={26},
  number={12},
  pages={123008},
  year={2024},
  publisher={IOP Publishing}
}

@article{dinelli2023non,
  title={Non-reciprocity across scales in active mixtures},
  author={Dinelli, Alberto and O’Byrne, J{\'e}r{\'e}my and Curatolo, Agnese and Zhao, Yongfeng and Sollich, Peter and Tailleur, Julien},
  journal={Nature Communications},
  volume={14},
  number={1},
  pages={7035},
  year={2023},
  publisher={Nature Publishing Group UK London}
}

@article{markovich2024nonreciprocity,
  title={Nonreciprocity and odd viscosity in chiral active fluids},
  author={Markovich, Tomer and Lubensky, Tom C},
  journal={Proceedings of the National Academy of Sciences},
  volume={121},
  number={19},
  pages={e2219385121},
  year={2024},
  publisher={National Academy of Sciences}
}

@article{gupta2022active,
  title={Active nonreciprocal attraction between motile particles in an elastic medium},
  author={Gupta, Rahul Kumar and Kant, Raushan and Soni, Harsh and Sood, AK and Ramaswamy, Sriram},
  journal={Physical Review E},
  volume={105},
  number={6},
  pages={064602},
  year={2022},
  publisher={APS}
}

@article{you2020nonreciprocity,
  title={Nonreciprocity as a generic route to traveling states},
  author={You, Zhihong and Baskaran, Aparna and Marchetti, M Cristina},
  journal={Proceedings of the National Academy of Sciences},
  volume={117},
  number={33},
  pages={19767--19772},
  year={2020},
  publisher={National Academy of Sciences}
}

@article{ivlev2015statistical,
  title={Statistical mechanics where Newton’s third law is broken},
  author={Ivlev, Alexei V and Bartnick, J{\"o}rg and Heinen, Marco and Du, C-R and Nosenko, V and L{\"o}wen, Hartmut},
  journal={Physical Review X},
  volume={5},
  number={1},
  pages={011035},
  year={2015},
  publisher={APS}
}

@article{te2023microscopic,
  title={From a microscopic inertial active matter model to the Schr{\"o}dinger equation},
  author={Te Vrugt, Michael and Frohoff-H{\"u}lsmann, Tobias and Heifetz, Eyal and Thiele, Uwe and Wittkowski, Raphael},
  journal={Nature Communications},
  volume={14},
  number={1},
  pages={1302},
  year={2023},
  publisher={Nature Publishing Group UK London}
}

@article{te2025metareview,
  title={Metareview: a survey of active matter reviews},
  author={Te Vrugt, Michael and Wittkowski, Raphael},
  journal={The European Physical Journal E},
  volume={48},
  number={3},
  pages={12},
  year={2025},
  publisher={Springer}
}

@article{menzel2013traveling,
  title={Traveling and resting crystals in active systems},
  author={Menzel, Andreas M and L{\"o}wen, Hartmut},
  journal={Physical review letters},
  volume={110},
  number={5},
  pages={055702},
  year={2013},
  publisher={APS}
}

@article{weis2025generalized,
  title={Generalized non-reciprocal phase transitions in multipopulation systems},
  author={Weis, Cheyne and Hanai, Ryo},
  journal={arXiv preprint arXiv:2507.16763},
  year={2025}
}

@article{yasuda2021nonreciprocality,
  title={Nonreciprocality of a micromachine driven by a catalytic chemical reaction},
  author={Yasuda, Kento and Komura, Shigeyuki},
  journal={Physical Review E},
  volume={103},
  number={6},
  pages={062113},
  year={2021},
  publisher={APS}
}

@article{kryuchkov2018dissipative,
  title={Dissipative phase transitions in systems with nonreciprocal effective interactions},
  author={Kryuchkov, Nikita P and Ivlev, Alexei V and Yurchenko, Stanislav O},
  journal={Soft Matter},
  volume={14},
  number={47},
  pages={9720--9729},
  year={2018},
  publisher={Royal Society of Chemistry}
}

@article{hosaka2021nonreciprocal,
  title={Nonreciprocal response of a two-dimensional fluid with odd viscosity},
  author={Hosaka, Yuto and Komura, Shigeyuki and Andelman, David},
  journal={Physical review E},
  volume={103},
  number={4},
  pages={042610},
  year={2021},
  publisher={APS}
}

@article{poncet2022soft,
  title={When soft crystals defy newton’s third law: Nonreciprocal mechanics and dislocation motility},
  author={Poncet, Alexis and Bartolo, Denis},
  journal={Physical Review Letters},
  volume={128},
  number={4},
  pages={048002},
  year={2022},
  publisher={APS}
}

@article{bratanov2015new,
  title={New class of turbulence in active fluids},
  author={Bratanov, Vasil and Jenko, Frank and Frey, Erwin},
  journal={Proceedings of the National Academy of Sciences},
  volume={112},
  number={49},
  pages={15048--15053},
  year={2015},
  publisher={National Academy of Sciences}
}

@article{thampi2016active,
  title={Active turbulence in active nematics},
  author={Thampi, SumeshP and Yeomans, JuliaM},
  journal={The European Physical Journal Special Topics},
  volume={225},
  number={4},
  pages={651--662},
  year={2016},
  publisher={Springer}
}

@article{dunkel2013minimal,
  title={Minimal continuum theories of structure formation in dense active fluids},
  author={Dunkel, J{\"o}rn and Heidenreich, Sebastian and B{\"a}r, Markus and Goldstein, Raymond E},
  journal={New Journal of Physics},
  volume={15},
  number={4},
  pages={045016},
  year={2013},
  publisher={IOP Publishing}
}

@article{wensink2012meso,
  title={Meso-scale turbulence in living fluids},
  author={Wensink, Henricus H and Dunkel, J{\"o}rn and Heidenreich, Sebastian and Drescher, Knut and Goldstein, Raymond E and L{\"o}wen, Hartmut and Yeomans, Julia M},
  journal={Proceedings of the national academy of sciences},
  volume={109},
  number={36},
  pages={14308--14313},
  year={2012},
  publisher={National Academy of Sciences}
}

@article{sokolov2007concentration,
  title={Concentration dependence of the collective dynamics of swimming bacteria},
  author={Sokolov, Andrey and Aranson, Igor S and Kessler, John O and Goldstein, Raymond E},
  journal={Physical review letters},
  volume={98},
  number={15},
  pages={158102},
  year={2007},
  publisher={APS}
}

@article{pandit2025particles,
  title={Particles and fields in minimal hydrodynamic models for active turbulence},
  author={Pandit, Rahul and Kiran, Kolluru Venkata},
  journal={Europhysics Letters},
  volume={150},
  number={1},
  pages={13001},
  year={2025},
  publisher={IOP Publishing}
}

@article{bowick2022symmetry,
  title={Symmetry, thermodynamics, and topology in active matter},
  author={Bowick, Mark J and Fakhri, Nikta and Marchetti, M Cristina and Ramaswamy, Sriram},
  journal={Physical Review X},
  volume={12},
  number={1},
  pages={010501},
  year={2022},
  publisher={APS}
}

@article{mandal2024molecular,
  title={A molecular origin of non-reciprocal interactions between interacting active catalysts},
  author={Mandal, Niladri Sekhar and Sen, Ayusman and Astumian, R Dean},
  journal={Chem},
  volume={10},
  number={4},
  pages={1147--1159},
  year={2024},
  publisher={Elsevier}
}

@article{nasouri2020exact,
  title={Exact phoretic interaction of two chemically active particles},
  author={Nasouri, Babak and Golestanian, Ramin},
  journal={Physical review letters},
  volume={124},
  number={16},
  pages={168003},
  year={2020},
  publisher={APS}
}

@article{liu2024self,
  title={Self-Organized Patterns in Non-Reciprocal Active Droplet Systems},
  author={Liu, Yutong and Kailasham, R and Moerman, Pepijn G and Khair, Aditya S and Zarzar, Lauren D},
  journal={Angewandte Chemie International Edition},
  volume={63},
  number={49},
  pages={e202409382},
  year={2024},
  publisher={Wiley Online Library}
}

@book{frisch1995turbulence,
  title={Turbulence: the legacy of AN Kolmogorov},
  author={Frisch, Uriel},
  year={1995},
  publisher={Cambridge university press}
}

@article{frisch1985singularity,
  title={On the singularity structure of fully developed turbulence},
  author={Frisch, U},
  journal={Turbulence and predictability in geophysical fluid dynamics and climate dynamics},
  year={1985}
}

@article{meneveau1991multifractal,
  title={The multifractal nature of turbulent energy dissipation},
  author={Meneveau, Charles and Sreenivasan, Katepalli R},
  journal={Journal of Fluid Mechanics},
  volume={224},
  pages={429--484},
  year={1991},
  publisher={Cambridge University Press}
}

@article{soto2014self,
  title={Self-assembly of catalytically active colloidal molecules: tailoring activity through surface chemistry},
  author={Soto, Rodrigo and Golestanian, Ramin},
  journal={Physical review letters},
  volume={112},
  number={6},
  pages={068301},
  year={2014},
  publisher={APS}
}

@article{saha2020scalar,
  title={Scalar active mixtures: The nonreciprocal Cahn-Hilliard model},
  author={Saha, Suropriya and Agudo-Canalejo, Jaime and Golestanian, Ramin},
  journal={Physical Review X},
  volume={10},
  number={4},
  pages={041009},
  year={2020},
  publisher={APS}
}

@article{guislain2024collective,
  title={Collective oscillations in a three-dimensional spin model with non-reciprocal interactions},
  author={Guislain, Laura and Bertin, Eric},
  journal={Journal of Statistical Mechanics: Theory and Experiment},
  volume={2024},
  number={9},
  pages={093210},
  year={2024},
  publisher={IOP Publishing}
}

@article{clerk2022introduction,
  title={Introduction to quantum non-reciprocal interactions: from non-Hermitian Hamiltonians to quantum master equations and quantum feedforward schemes},
  author={Clerk, Aashish},
  journal={SciPost Physics Lecture Notes},
  pages={044},
  year={2022}
}

@article{sahoo2025nonreciprocal,
  title={Nonreciprocal Model B: The role of mobilities and nonreciprocal interfacial forces},
  author={Sahoo, Bibhut and Mandal, Rituparno and Sollich, Peter},
  journal={arXiv preprint arXiv:2508.02814},
  year={2025}
}

@article{al2025non,
  title={Non-Reciprocity, Metastability, and Dynamic Reconfiguration in Co-Assembly of Active and Passive Particles},
  author={Al Harraq, Ahmed and Patel, Ruchi and Lee, Jin Gyun and Owoyele, Ope and Chun, Jaehun and Bharti, Bhuvnesh},
  journal={Advanced Science},
  volume={12},
  number={4},
  pages={2409489},
  year={2025},
  publisher={Wiley Online Library}
}

@article{klamser2025directed,
  title={Directed percolation transition to active turbulence driven by non-reciprocal forces},
  author={Klamser, Juliane U and Berthier, Ludovic},
  journal={arXiv preprint arXiv:2510.04575},
  year={2025}
}

@article{ryu2023dynamics,
  title={Dynamics in non-Hermitian systems with nonreciprocal coupling},
  author={Ryu, Jung-Wan},
  journal={Physical Review A},
  volume={108},
  number={5},
  pages={052205},
  year={2023},
  publisher={APS}
}

@article{montbrio2018kuramoto,
  title={Kuramoto model for excitation-inhibition-based oscillations},
  author={Montbri{\'o}, Ernest and Paz{\'o}, Diego},
  journal={Physical review letters},
  volume={120},
  number={24},
  pages={244101},
  year={2018},
  publisher={APS}
}

@article{sompolinsky1986temporal,
  title={Temporal association in asymmetric neural networks},
  author={Sompolinsky, Haim and Kanter, Ido},
  journal={Physical review letters},
  volume={57},
  number={22},
  pages={2861},
  year={1986},
  publisher={APS}
}

@article{hong2011kuramoto,
  title={Kuramoto model of coupled oscillators with positive and negative coupling parameters: an example of conformist and contrarian oscillators},
  author={Hong, Hyunsuk and Strogatz, Steven H},
  journal={Physical Review Letters},
  volume={106},
  number={5},
  pages={054102},
  year={2011},
  publisher={APS}
}

@article{reisenbauer2024non,
  title={Non-Hermitian dynamics and non-reciprocity of optically coupled nanoparticles},
  author={Reisenbauer, Manuel and Rudolph, Henning and Egyed, Livia and Hornberger, Klaus and Zasedatelev, Anton V and Abuzarli, Murad and Stickler, Benjamin A and Deli{\'c}, Uro{\v{s}}},
  journal={Nature Physics},
  volume={20},
  number={10},
  pages={1629--1635},
  year={2024},
  publisher={Nature Publishing Group UK London}
}

@article{garces2025phase,
  title={Phase transitions in single species Ising models with non-reciprocal couplings},
  author={Garc{\'e}s, Adri{\`a} and Levis, Demian},
  journal={Journal of Statistical Mechanics: Theory and Experiment},
  volume={2025},
  number={4},
  pages={043205},
  year={2025},
  publisher={IOP Publishing}
}

@article{avni2025nonreciprocal,
  title={Nonreciprocal ising model},
  author={Avni, Yael and Fruchart, Michel and Martin, David and Seara, Daniel and Vitelli, Vincenzo},
  journal={Physical Review Letters},
  volume={134},
  number={11},
  pages={117103},
  year={2025},
  publisher={APS}
}

@article{alert2022active,
  title={Active turbulence},
  author={Alert, Ricard and Casademunt, Jaume and Joanny, Jean-Fran{\c{c}}ois},
  journal={Annual Review of Condensed Matter Physics},
  volume={13},
  number={1},
  pages={143--170},
  year={2022},
  publisher={Annual Reviews}
}
\end{document}